\documentstyle[prl,aps,multicol]{revtex}

\newcommand{\tS}{\widetilde \Sigma}

\newcommand{\wh}{\widetilde h}

\newcommand{\tD}{\widetilde \Delta}

\newcommand{\re}{{{\rm I} \!  {\rm R}}}
\newcommand{\ro}{{ r_0}}

\newcommand{\zz}{{\bf \rm Z}}
\newcommand{\cW}{{\cal W}}

\begin{document}
\vskip 0 true cm
\flushbottom
\title{\bf A uniqueness theorem for the AdS soliton}

\author{G.J.\ Galloway$^{\ast \,a}$, S.\ Surya$^{\dag \, \ddag \,b}$,  E.\
Woolgar$^{\ddag \,c}$}
\address{
$^\ast$ Dept.\ of Mathematics, University of
Miami, Coral Gables, FL 33124, USA. \\
$^\dag$ Dept.\ of Physics and Astronomy, University of British
Columbia, Vancouver, BC, Canada V6T 1Z1. \\
$^\ddag$ Dept.\ of Mathematical Sciences, University of Alberta,
Edmonton, AB, Canada T6G 2G1.
}
\date{\today}
\maketitle
\begin{abstract}
The stability of physical systems depends on the existence of a
state of least energy. In gravity, this is guaranteed by the
positive energy theorem. For topological reasons this fails for
nonsupersymmetric Kaluza-Klein compactifications, which can decay
to arbitrarily negative energy. For related reasons, this also
fails for the AdS soliton, a globally static, asymptotically
toroidal $\Lambda<0$ spacetime with negative mass. Nonetheless,
arguing from the AdS/CFT correspondence, Horowitz and Myers
proposed a new positive energy conjecture, which asserts that the
AdS soliton is the unique state of least energy in its asymptotic
class. We give a new structure theorem for static $\Lambda<0$
spacetimes and use it to prove uniqueness of the AdS soliton. Our
results offer significant support for the new positive energy
conjecture and add to the body of rigorous results inspired by the
AdS/CFT correspondence.
\end{abstract}

\begin{multicols}{2}

\noindent
The positive energy theorem \cite{schyau,witten81} singles out
Minkowski spacetime as the ``ground state'', or spacetime of lowest
mass-energy, within the class of asymptotically flat spacetimes with
local energy density $\ge 0$ and without naked singularities. In the
presence of a negative cosmological constant the appropriate ground
state is anti-de Sitter spacetime \cite{ghw83}.  These ground
states are regular, globally static, supersymmetric, and of constant
curvature.  Moreover, it is known that Minkowski
spacetime is the unique asymptotically flat, regular, stationary vacuum
spacetime \cite{lich}. The analogous uniqueness result for the
asymptotically globally adS case is proved in \cite{bgh,CS}.

A simple scaling argument suggests that a ground state cannot have
negative mass, for then it could be scaled to produce a state of
even lesser mass \cite{bubble}. Consider, then, the surprising
properties of the adS soliton, first examined by Horowitz and
Myers \cite{hm}, which is a negative mass, globally static
Einstein spacetime with cosmological constant $\Lambda <0$. The
metric in $n+1\geq 4$ spacetime dimensions is
\begin{equation}
ds^2 = -r^2dt^2 + \frac{1}{V(r)}dr^2 + V(r)d\phi^2
+r^2 \sum\limits_{i=1}^{n-2}(dy^i)^2  \label{Intro1}
\end{equation}
where $V(r)= \frac{r^2}{\ell^2} \left ( 1-\frac{\ro^{n}}{r^{n}}
\right )$, $\ell^2=-\frac{n(n-1)}{2\Lambda}$, and $\ro$ is a
constant. Regularity demands that $\phi$ be identified with period
$\beta_0=\frac{4\pi\ell^2}{n\ro}$. The periods of the $y^i$ are
arbitrary. The soliton is ``asymptotically locally anti-de
Sitter'' with boundary at conformal infinity (scri) foliated by
spacelike $(n-1)$-tori. The time slices of spacetime itself, when
conformally completed, are topologically the product of an $(n-2)$-torus
and a disk (a solid torus in $3+1$). The soliton spacetime is
neither supersymmetric nor of constant curvature, but has minimal
energy under small metric perturbations \cite{hm,cm}. Remarkably,
one cannot vary the soliton mass by simple scaling. To see why,
note that rescaling the parameter $r_0\to kr_0$ in (\ref{Intro1})
has the same effect as the coordinate transformation
$(t,r,y^i,\phi) \mapsto (kt,k^{-1}r,ky^i,k\phi)$. Thus the new
metric (with parameter $kr_0$) is {\it isometric} to the original
one, provided the conformal boundary data are chosen to agree,
so they must have the same physical mass.

Horowitz and Myers found that the negative mass of the adS soliton
has a natural interpretation as the Casimir energy of a
non-supersymmetric gauge theory on the conformal boundary. If a
non-supersymmetric version of the adS/CFT conjecture holds
\cite{adscft}, as is generally hoped, then this would indicate
that the soliton is the lowest energy solution with these boundary
conditions. This led them to postulate a new positive energy
conjecture, that the soliton is the unique lowest mass solution
for all spacetimes in its asymptotic class.  The validity of this
conjecture is thus an important test of the non-supersymmetric
version of the adS/CFT correspondence. The conjecture is all the
more remarkable because the soliton topology has certain circles
that are not contractible at infinity but are contractible in the
bulk. This leads to the failure of spinorial methods to produce a
positive energy theorem here and is linked to a known instability
in Kaluza-Klein theory\cite{bubble,BP}.

As support for the new positive energy conjecture, we will give a
uniqueness theorem for the adS soliton, singling it out as the
only suitable ground state in the class of spacetimes with similar
asymptotics. Our theorem is similar in spirit to \cite{lich,bgh},
but relates to asymptotically ``locally'' adS spacetimes with
Ricci flat conformal boundary. The proof is based on the fact that
the soliton indeed shares one important geometric property with
known ground states: its universal cover admits a foliation by
totally geodesic null surfaces ruled by complete, achronal null
geodesics called {\it null lines}. In other words, the spacetime
admits non-focusing plane waves. A similar idea underlies the
approach to mass positivity in \cite{psw}, a significant
difference being that the null lines we will construct here do not
approach scri. Our construction of the null surfaces relies on
results in \cite{Gal}.

Our arguments will be of necessity terse. Herein we give the
flavour of the proof; details will appear in \cite{us}. Our results
considerably generalize a uniqueness result of \cite{KP}. Of related
interest is a uniqueness theorem of Anderson for asymptotically
hyperbolic Einstein metrics on $4$-dimensional Riemannian manifolds
\cite{Anderson}.

Following the formalism  of Chru\'sciel and Simon \cite{CS}, we consider
static spacetimes of the form
\begin{equation}
M^{n+1} = \re\times \Sigma, \qquad g= -N^2dt^2 \oplus h \quad ,
\label{FE1}\end{equation} where $(\Sigma,h,N)$ is
\emph{conformally compactifiable}. Thus we assume that $\Sigma$ is
the interior of a compact manifold with boundary $\widetilde
\Sigma= \Sigma\cup \partial \widetilde \Sigma$ such that $\,$ (a)
$1/N$ extends to a smooth function $\widetilde N$ on $\widetilde
\Sigma$, with $\widetilde N=0$ and $d\widetilde N\ne 0$ along
$\partial \widetilde \Sigma$ and $\,$ (b) $N^{-2}h$ extends  to a
smooth Riemannian metric $\widetilde h$. $(M,g)$ conformally
embeds into $(\re \times \tS, -dt^2+\wh)$ and hence admits a
natural timelike scri structure, making precise, in the static
setting, what is meant by ``asymptotically locally adS''.

We assume that $(\Sigma,h,N)$ obeys the static vacuum field
equations
\begin{eqnarray}
R_{ab}& = & N^{-1}\nabla_a\nabla_b N +\frac{2\Lambda}{n-1} h_{ab}
\label{FE2}\,  , \\
\Delta N & = & -\frac{2\Lambda}{n-1} N \, ,
\label{FE3}
\end{eqnarray}
where $\Delta=\nabla^2$, $\nabla_a$ is the covariant derivative on
$(\Sigma, h_{ab})$ and $R_{ab}$ is its Ricci tensor. In
terms of the rescaled metric $\widetilde h$, and associated $\tD, \widetilde
\nabla_a$ and $\widetilde R_{ab}$,
\begin{eqnarray}
\widetilde R_{ab}&=&\frac{-(n-1)}{\widetilde N}\widetilde \nabla_a\widetilde
\nabla_b \widetilde N,\label{FE4}\\
\widetilde N\,\widetilde \Delta \widetilde
N&=&\frac{2\Lambda}{n-1}+n\widetilde W ,\label{FE5}
\end{eqnarray}
where $\widetilde W:={\widetilde h^{ab}}\widetilde \nabla_a\widetilde N\widetilde
\nabla_b\widetilde N =N^{-2}h^{ab}\nabla_a N\nabla_b
N\label{FE6}$.

Recall that the rescaled metric $\widetilde h = N^{-2}h$,
sometimes called the \emph{Fermat (optical) metric}, has physical
significance: geodesics of $(\tS,\widetilde h)$ correspond to the
spatial paths of light rays in $(M,g)$.  That is, null geodesics
in $(M,g)$ project in the obvious way to geodesics in
$(\tS,\widetilde h)$ (when suitably parametrized). Conversely, a
geodesic $\gamma$ in $\tS$ (viewed as the slice $t=0$) passing
through a point $p\in \tS$ lifts to a unique future directed null
geodesic $\eta$ passing through $p$.  Moreover, if $\gamma$ is a
length minimizing geodesic segment in $(\tS,\widetilde h)$ then it
lifts to an \emph{achronal} null geodesic segment $\eta$.  Thus, a
line (inextendible geodesic, length minimizing on each segment) in
$(\tS,\widetilde h)$ lifts to a null line (achronal inextendible
null geodesic) in $(M,g)$.  This basic fact is used in an
essential way in our arguments.

Our uniqueness result (Theorem 2) for the adS soliton is obtained
as a consequence of a more general structure result (Theorem 1)
which assumes a certain \emph{convexity condition} near infinity.
As discussed below, this convexity condition is related to the
sign of the mass (or total energy) of the spacetime. In general,
as follows from eq.\ (\ref{FE4}) and the $C^2$ smoothness of
$\widetilde h_{ab}$ at $\widetilde N=0$, the conformal boundary
$\partial\widetilde \Sigma=\{\widetilde N=0\}$ is totally geodesic
in $(\widetilde \Sigma,\widetilde h)$. We say that $(\Sigma,h,N)$
\emph{satisfies condition (C)} if there exists a neighbourhood of
scri in which each level surface $\widetilde N=c$ is \emph{weakly
convex} in $(\tS,\wh)$, i.e., the second fundamental form with
respect to the outward normal of the level surface is positive
semi-definite.  Equivalently, this condition requires the
principal curvatures of each level surface sufficiently close to
scri be non-negative.

\noindent {\bf Theorem 1:} {\sl Consider a static spacetime as in
(\ref{FE1}) such that ({\it i}) $(\Sigma,h,N)$ is conformally
compactifiable, (\it {ii}) the static vacuum field equations hold,
and ({\it iii}) condition (C) holds. Then the Riemannian universal
cover (${\widetilde \Sigma}^\ast, {\widetilde h}^\ast)$ of
($\widetilde \Sigma,\widetilde h)$ splits isometrically as
\begin{equation}
{\widetilde \Sigma}^\ast =  \re^k\times \cW, \qquad
{\widetilde h}^\ast = h_E\oplus \sigma\,\label{ST1}
\end{equation}
where $(\re^k, h_E)$ is standard $k$-dimensional Euclidean
space and $(\cW,\sigma)$ is a \emph{compact} simply connected Riemannian
manifold-with-boundary. The Riemannian universal cover
$(\Sigma^\ast,h^\ast)$ of $(\Sigma,h)$ splits
isometrically as a warped product of the form,
\begin{equation}
\Sigma^\ast =  \re^k\times \cW_0, \qquad  h^\ast = (N^{\ast 2}
h_E)\oplus \sigma_0 \, ,\label{ST2}
\end{equation}
where  $N^\ast=N\circ \pi$ ($\pi =$ covering map) depends
only on $\cW_0$, and $(\cW_0,\sigma_0)$ is a complete simply connected
Riemannian manifold  such that $(\cW_0,\sigma_0,N)$ is conformally
compactifiable. }

Theorem 1 is similar in spirit to a result of Cheeger and Gromoll
\cite{CG} concerning the structure of compact Riemannian manifolds
of non-negative Ricci curvature. It implies a strong structure
result for the fundamental group $\Pi_1(\widetilde \Sigma)$, cf.\
\cite{CG}.

Let us now consider how condition (C) relates to the sign of the mass.
As we are using the conformal approach, the Ashtekar-Magnon
\cite{AM} mass expression involving the electric part of the Weyl tensor
is especially convenient.  Consider a conformally compactifiable
static spacetime $(\Sigma,h,N)$, and view $\Sigma$ as the slice $t=0$.
Setting $x =\widetilde N$, the metric $\widetilde h$ near the conformal
boundary $x = 0$ may be written as
\begin{equation}
\widetilde h = \widetilde W^{-1}dx^2 + h_{AB}(x,x^C)dx^Adx^B \, ,
\label{gauge}
\end{equation}
where $h_{AB}$ is the induced metric on $x=const$ surfaces. Let
$T^a$ denote the future-timelike unit normal to $\widetilde
\Sigma$ in the rescaled spacetime metric, and on $\widetilde
\Sigma$, let $n^a = -{\sqrt{\widetilde W}}\partial_x$ be the
outward pointing unit normal field to the slices $x= const \,$
near the conformal boundary.  The Weyl mass is then given, up to a
positive constant, by $\int_{\partial\widetilde \Sigma}\mu
\widetilde {dA}$, where $\widetilde {dA}$ is the volume element on
$\partial\widetilde \Sigma$. The \emph{mass aspect} $\mu$, up to a
positive constant, is given by
\begin{equation}
\mu = \lim_{x\to 0}\frac{\widetilde E_{ac}T^aT^c}{x^{n-2}}
= \lim_{x\to 0}\frac{\widetilde C_{abcd}n^bn^dT^aT^c}{x^{n-2}} \, ,
\end{equation}
where ${\widetilde C_{abcd}}$ is the Weyl tensor (and ${\widetilde
E_{ac}}$ its electric part) of the conformal spacetime metric
$\widetilde g= N^{-2} g$.

In the static setting, the mass aspect $\mu$ can be directly
related to the geometry of $(\widetilde \Sigma,\widetilde h)$.
One finds, using the field equations, that, up to a positive constant,
$\mu = -\partial^{n-2} \widetilde R/\partial x^{n-2}|_{x=0}$.
(Chru\'sciel and Simon \cite{CS} had previously identified, in the $3+1$
dimensional case,
$-\partial \widetilde R/\partial x|_{x=0}$ as the mass aspect.)

Let $\widetilde H$ denote the mean curvature function of the
slices $x = const \,$ with respect to the outward normal $n^a$;
along each such slice, $\widetilde H = \widetilde \nabla_a n^a =$
the trace of the second fundamental form $=$ the sum of the
principal curvatures.  Using the field equations and the Gauss
equation, one can show that $\widetilde H$ is related to $\mu$ by
\begin{equation}
(n-2)\sqrt{\widetilde W}\,\widetilde H = -\frac{x^{n-1}}{2(n-1)}
\mu +{\cal O}(x^n)\label{AG7}
\end{equation}
when the conformal boundary has Ricci flat induced metric. In
order to establish Equation (\ref{AG7}), we must carry out a
Fefferman-Graham type boundary analysis \cite{FG}, for the metric
(\ref{gauge}) relevant to our situation, and subject to the field
equations (\ref{FE4}) and (\ref{FE5}). This analysis implies
$\widetilde R|_{x=0} =0$ and $\partial^\ell\widetilde R/\partial
x^{\ell}|_{x=0} = 0$ for $1\le\ell \le n-3$.

For Ricci flat conformal boundary, as is the case for the soliton,
equation (\ref{AG7}) implies that if the mass aspect $\mu$ is
(pointwise) negative, the level surfaces $\widetilde N = c$ near
conformal infinity are outwardly \emph{mean convex} in $(\tS,
\wh)$, i.e., have strictly positive mean curvature (and hence the
sum of the principal curvatures is positive). In other words, if
the mass aspect is negative (as it is for the adS soliton) then
condition (C) holds \emph{in the mean}. However, our proof of
uniqueness of the adS soliton requires not just mean convexity,
but (weak) convexity near infinity, and so we need to impose the
following condition.  We say that $(\Sigma,h,N)$ \emph{satisfies
condition (S)} provided the principal curvatures of the level
surfaces $\widetilde N =c$ near infinity are either all
non-negative or all non-positive.  This condition holds trivially
in the adS soliton, since all but one of the principal curvatures
vanish.  It also holds in the Kottler spacetimes, regardless of
the sign of the mass.

\noindent {\bf Theorem 2:} {\sl Consider a static spacetime as in
(\ref{FE1}) such that {\rm (i)} $(\Sigma,h,N)$ is conformally
compactifiable, {\rm (ii)} the static vacuum field equations hold,
and {\rm (iii)} condition (S) holds. Suppose in addition,  that
\begin{enumerate}
\item[\rm (a)] The boundary geometry of $(\widetilde \Sigma,\widetilde h)$
is the same as that of (\ref{Intro1}), i.e., $\partial
\widetilde \Sigma = T^{n-2}\times S^1$, $\widetilde h|_{\partial \widetilde
\Sigma} = d\phi^2+ \sum\limits_{i=1}^{n-2}(dy^i)^2 $, with the
same periods for $\phi$ and the $y^i$.
\item[\rm (b)] The mass aspect $\mu$ of $(\Sigma,h,N)$ is pointwise
negative.
\item[\rm (c)]
Given the inclusion map $i:\partial\tS \rightarrow \tS$, the kernel of the
induced homomorphism of fundamental groups, $i_*:\Pi_1(\partial\tS) \rightarrow \Pi_1(\tS)$, is
generated by the $S^1$ factor.
\end{enumerate}
\vspace{-.1in}
Then the spacetime (\ref{FE1}) determined by $(\Sigma,h,N)$ is
isometric to the adS  soliton~(\ref{Intro1}). }

Assumption (a) is a natural boundary condition. Assumption (b),
together with condition (S), guarantees that condition (C) of
Theorem 1 holds.
Assumption (c) asserts that the generator of the $S^1$
factor  is contractible in
$\tS$, and moreover, that any loop in $\partial\tS$ contractible
in $\tS$ is a multiple of the generator.  As discussed in the
proof, assumptions (a) and (c)
together imply that $\Pi_1(\tS)\approx \zz^{n-2}$.  Were we
to adopt the latter condition in lieu of assumption (c),
then one could only conclude that $(\Sigma,h,N)$ is
\emph{locally} isometric to the adS soliton (the universal covers
will be isometric, however). For further discussion of this
\emph{discrete} nonuniqueness relevant to the adS soliton in $3+1$
dimensions, see \cite{Anderson}. We note that in $3+1$ dimensions,
assumption (b) (when $\mu$ is constant), and the condition
$\Pi_1(\tS)\approx \zz^{n-2}$, hold automatically, cf.\ \cite{CS,GSWW}.

We now sketch the proofs of Theorems 1 and 2; details
will appear in \cite{us}.

\noindent
\emph{Sketch of the proof of Theorem 1.}  We proceed inductively,
working in $({\widetilde \Sigma}^\ast, {\widetilde h}^\ast)$. If
${\widetilde \Sigma}^\ast$ is compact then Theorem 1 holds with
$k=0$. So suppose ${\widetilde \Sigma}^\ast$ is noncompact. In this
case there is a procedure for constructing a line in
$({\widetilde \Sigma}^\ast, {\widetilde h}^\ast)$.
Fix a point $p$ in the interior, and let $\{p_i\}$ be a sequence of
points uniformly bounded away from $\partial{\widetilde
\Sigma}^\ast$, such that the distance from $p$ to $p_i$ tends to infinity.
For each $i$, $p$ and $p_i$
can be joined by a length minimizing geodesic segment $\gamma_i$ which
cannot meet the boundary (since it's totally geodesic).
In fact, by condition (C), the segments $\gamma_i$
must be uniformly bounded away from $\partial{\widetilde
\Sigma}^\ast$. Each geodesic segment $\gamma_i$ will have a midpoint
$r_i$.  Now since $\widetilde \Sigma$ is compact, it will have a
compact fundamental domain $D$ in ${\widetilde \Sigma}^\ast$. For
each point in the covering space, there will be a covering space
transformation mapping that point to a point in $D$. We therefore
apply to each geodesic segment $\gamma_i$ a covering space
transformation that maps $r_i$ into $D$. This produces a sequence of
minimizing geodesic segments $\sigma_i$, still uniformly bounded away
from $\partial{\widetilde \Sigma}^\ast$, whose lengths are
unbounded in both directions. By standard
compactness results, this sequence
possesses a limit curve which is a complete, length minimizing
geodesic, i.e., a line, in the interior of ${\widetilde \Sigma}^\ast$.

By the relationship between Fermat and null geodesics, this line lifts
to a complete null line $\eta$ in the physical covering spacetime
$(M^\ast,g^\ast)$, which is null geodesically complete and
obeys the null energy condition, $R_{ab}X^aX^b \ge 0$ for all null
vectors $X^a$. Then a null line will exist only under
special circumstances.  As proved in \cite{Gal}, $\eta$ must be
contained in a smooth achronal edgeless null hypersurface $\cal H$
which is \emph{totally geodesic} (i.e., has vanishing expansion and
shear).  In a static spacetime this has further consequences.  Since
$\Sigma^\ast$ (viewed as the slice $t=0$) is totally geodesic,
$\cal H$ meets $\Sigma^\ast$ in a totally geodesic submanifold
$\cal W$ of codimension one in $\Sigma^\ast$.  But by moving $\cal
H$ invariantly under the flow generated by $\partial/\partial t$, we
see that spacetime is actually foliated by totally geodesic null
hypersurfaces, and this gives rise to a foliation $\{{\cal W}_u\}$ of
$\Sigma^\ast$ by totally geodesic hypersurfaces in
$\Sigma^\ast$.  Moreover, it can be seen that this foliation is
achieved by exponentiating out along the unit speed normal geodesics
to $\cal W$ in the \emph{Fermat} metric.  The physical metric
then takes the form,
\begin{equation}
h^\ast = {N}^{\ast 2}(u,x^A)du^2 + h_{AB}(x)dx^Adx^B \, .\label{warped}
\end{equation}
Up to this point we have only used the null energy condition.  Now
using the field equations in a more explicit way, one can show
$\partial {N}^\ast/\partial u = 0$, and hence the metric
(\ref{warped}) is a genuine warped product.  By multiplying
(\ref{warped}) by $({N}^\ast)^{-2}$, and showing that everything
extends smoothly to the boundary, we conclude that
$({\widetilde \Sigma}^\ast, {\widetilde h}^\ast)$
is isometric to the Riemannian product $(\re\times {\cal W}, du^2 +
d\widetilde\sigma^2)$, where $({\cal W}, d\widetilde\sigma^2)$ is a
complete Riemannian manifold-with-boundary.  If $\cal W$ is compact
then Theorem 1 holds with $k=1$. If $\cal W$ is noncompact then one
can carry out essentially the same procedure again to construct a line
in $\cal W$, lift it to a new null line spatially orthogonal to the
first, and split off another $\re$ factor.  One can continue splitting
off $\re$ factors until what remains is compact.

\noindent \emph{Sketch of the proof of Theorem 2.}  Let
$(\Sigma_0,h_0, N_0)$ denote the adS soliton associated with the
boundary data in assumption (a), and let
$(\widetilde\Sigma_0,\widetilde h_0, \widetilde N_0)$ denote the
corresponding conformally compactified soliton. Assumption (b),
condition (S) and equation (\ref{AG7}) imply that condition (C)
holds, and hence we can apply Theorem 1. As shown below,
$\Pi_1(\tS) \approx \zz^{n-2}$, from which it follows that $k
=n-2$.  Thus, the universal cover ${\widetilde\Sigma}^\ast$ splits
isometrically as $\re^{n-2}\times \cW$, where $\cW$ is
diffeomorphic to a disk, and the metric on $\cW$ is determined by
the field equations (\ref{FE4}), (\ref{FE5}). From topological
censorship \cite{GSWW}, it follows that the homomorphism $i_* :
\Pi_1(\partial\widetilde \Sigma) \to \Pi_1(\widetilde \Sigma)$ is
onto.  But $\partial\widetilde \Sigma = A\times B$, where $A$ is
the $(n-2)$ torus and $B$ is the circle of assumption (a), whence
$\Pi_1(\partial \Sigma)\approx \Pi_1(A)\times\Pi_1(B)$. Since by
assumption~(c), $\ker i_* = \Pi_1(B)$, it follows  that
$i_*|_{\Pi_1(A)}:\Pi_1(A)\to \Pi_1(\widetilde \Sigma)$ is an
isomorphism. This implies that the covering transformations of
${\widetilde\Sigma}^\ast$ are in one-to-one correspondence, via
$i_*|_{\Pi_1(A)}$, to those of $A^\ast$, the universal covering
space of $A$. Thus, $\widetilde \Sigma \simeq {\widetilde
\Sigma}^\ast /\Pi_1(\widetilde \Sigma) \simeq (\re^{n-2} \times
\cW)/\Pi_1 (A) \simeq ({A}^\ast/\Pi_1(A))\times \cW \simeq A\times
\cW$, i.e., $\widetilde \Sigma$ is isometric to $T^{n-2}\times
\cW$, where $T^{n-2}$ is the torus in assumption~(a).  Because of
the product structure of $\widetilde\Sigma$ and the fact that
$\widetilde N$ depends only on $\cW$, the field equations
(\ref{FE4}) and (\ref{FE5}) descend to the disk $\cW$, and can be
solved explicitly and uniquely, subject to the appropriate
boundary conditions on $\partial\widetilde\Sigma$.  The result is
that $\widetilde\Sigma$ is isometric to $\widetilde\Sigma_0$ and
$\widetilde N = \widetilde N_0$, from which we conclude that
$(\Sigma, h, N)$ is isometric to the adS soliton $(\Sigma_0, h_0,
N_0)$.

\noindent {\bf Acknowledgements:} This work was partially
supported by grants from the NSF (DMS-9803566, DMS-0104042) and
NSERC. SS was supported by a Pacific Institute (PIms)
post-doctoral fellowship. GJG wishes to thank DAMTP, University of
Cambridge, for hospitality during the completion of part of this
work. \vskip 0.2cm

\small{
\noindent $^a$galloway@math.miami.edu. \\
\noindent $^b$ssurya@pims.math.ca. \\
\noindent $^c$ewoolgar@math.ualberta.ca.}

\end{multicols}
\end{document}